\documentclass{aa}     
\usepackage{graphics}
\usepackage{psfig}
\usepackage{latexsym}
\begin{document}

\newcommand{\zs} {$\zeta_\star$}
\newcommand{\vsini} {$v\,\sin i$}
\newcommand{\Sc} {Sect.}
\newcommand{\cmc} {cm$^{-3}$}
\newcommand{\cmq} {cm$^{-2}$}
\newcommand{\csqg} {cm$^2$ g$^{-1}$}
\newcommand{\kms} {km\,s$^{-1}$}
\newcommand{\myr} {$M_\odot$ yr$^{-1}$}
\newcommand{\Myr} {$M_\odot$ yr$^{-1}$}
\newcommand{\um} {$\mu$m}
\newcommand{\mic} {$\mu$m}
\newcommand{\sbr} { erg cm$^{-2}$ s$^{-1}$ sr$^{-1}$}
\newcommand{\sbu} { erg cm$^{-2}$ s$^{-1}$ sr$^{-1}$}
\newcommand{\Lsun} {L$_\odot$}
\newcommand{\Msun} {M$_\odot$}
\newcommand{\MJ} {M$_{\rm J}$}
\newcommand{\Teff} {T$_{\rm{eff}}$}
\newcommand{\Tstar} {T$_{\rm{eff}}$}
\newcommand{\Lstar} {L$_\star$}
\newcommand{\Rstar} {R$_\star$}
\newcommand{\Mstar} {M$_\star$}
\newcommand{\Md} {M$_{\rm D}$}
\newcommand{\Rd} {R$_{\rm D}$}
\newcommand{\Rin} {R$_i$}
\newcommand{\pms} {pre-main--sequence}
\newcommand{\DV} {$\Delta V$}
\newcommand{\Dm} {$\Delta m$}
\newcommand{{\AV}} {A$_{\rm V}$}
\newcommand{{\AJ}} {A$_{\rm J}$}
\newcommand{{\cii}} {$^{13}$CO(2--1)}
\newcommand{{\ciii}} {C$^{18}$O(2--1)}

\newcommand{\simless}{\mathbin{\lower 3pt\hbox
      {$\rlap{\raise 5pt\hbox{$\char'074$}}\mathchar"7218$}}} 
\newcommand{\simgreat}{\mathbin{\lower 3pt\hbox
     {$\rlap{\raise 5pt\hbox{$\char'076$}}\mathchar"7218$}}} 

\title{The kinematic relationship between disk and jet in the DG~Tauri system}
 
\author {Leonardo Testi\inst{1}, Francesca Bacciotti\inst{1},
Anneila I. Sargent\inst{2}, Thomas P. Ray\inst{3},
Jochen Eisl\"{o}ffel\inst{4}
}
%
\institute{
    Osservatorio Astrofisico di Arcetri, INAF, Largo E.Fermi 5,
    I-50125 Firenze, Italy
\and
    California Institute of Technology, MS~105-24, Pasadena, CA~91125, USA
\and
    Dublin Institute for Advanced Studies, 5 Merrion Square Dublin 2, Ireland
\and
    Th\"uringer Landessternwarte Tautenburg, Sternwarte 5,
    D-07778 Tautenburg, Germany
}

\offprints{lt@arcetri.astro.it}
\date{Received ...; accepted ...}
 
\titlerunning{The DG~Tauri disk-jet system}
\authorrunning{Testi et al.}
 
\begin{abstract}
{
We present high angular resolution millimeter wavelength continuum
and \cii\ observations of the circumstellar disk 
surrounding the T\,Tauri star DG~Tauri. We show that
the velocity pattern in the inner regions of the disk is consistent
with Keplerian rotation about a central 0.67\,M$_\odot$ star.
The disk rotation is also consistent
with the toroidal velocity pattern in the initial channel of the optical 
jet, as inferred from HST spectra of the first de-projected 100~AU from the 
source. Our observations support the tight relationship between disk
and jet 
kinematics postulated by the popular magneto-centrifugal models for jet 
formation and collimation.
}
\end{abstract}

\maketitle

\section{Introduction}

The interplay between accretion and ejection of matter is believed to be a
crucial element in the formation of stars. In particular, stellar jets
may contribute substantially to the removal of excess angular momentum from
the system, thereby allowing the central star to accrete to its final mass
(e.g. Eisl\"{o}ffel et al.~\cite{Eea00},
K\"{o}nigl \& Pudritz~\cite{KP00}, Shu et
al.~\cite{Sea00}). Most of the proposed models
invoke the simultaneous action of magnetic
and centrifugal forces in a rotating star/disk system threaded to open magnetic
field lines. Even if widely accepted, these models have not yet been tested
observationally on the launching scale
(a few AU from the star), although this may be possible with the coming
generation of interferometers.
%
   On much larger scales, there is some evidence
   for the requisite relationship between the jet
   and envelope kinematics in the HH\,212 protostellar
   system (Davis et al.~\cite{Dea00}). Hints of rotation
   are seen in the H$_2$ jet knots at 2$\times$10$^3$
   to 10$^4$\,AU from
   the powering source; the sense of rotation is the
   same as that of the flattened envelope detected in
   NH$_3$ VLA observations (Wiseman et al.~\cite{Wea01}).
   Although encouraging, these measurements probe regions too far from
   the central source to allow detailed comparison between the
   disk and jet kinematics.

Protostellar disk/jet systems are too embedded 
to probe the jet close to the launching region 
with current techniques which rely on optical and near infrared
observations. Moreover, the kinematics of
disk/envelope systems may encompass both rotational and infall motions,
hampering tests of disk-jet interaction models.
Optically visible T\,Tauri stars, which have associated disks but little
remnant envelopes, are much more suitable candidates for such studies.

 The optical jet from the T\,Tauri star DG\,Tauri
 has been extensively studied at high resolution
 in recent years and displays properties that are in general agreement
 with magneto-centrifugal models for jet-launching (Dougados et
 al.~\cite{Douea00}; Bacciotti et al.~\cite{Bea00}; \cite{Bea02}).
The latter
studies showed that the flow appears to have an onion-like kinematic
structure, with the faster and more collimated flow continuously bracketed in
a wider and slower one.  The flow becomes gradually denser and more excited
from the edges toward the axis. The mass loss rate in the flow
is about one tenth of
the estimated mass accretion rate through the disk (Bacciotti et
al.~\cite{Bea00}). 
Even more interestingly, for the spatially
resolved flow component at moderate velocity (peaked at $-$70\,\kms) 
systematic offsets in the radial velocity of the lines have been found in pairs
of slits symmetrically located with respect to the jet axis (Bacciotti et
al.~\cite{Bea02}). If these results are interpreted as rotation, then
the jet is rotating clockwise (looking toward the source) with average
toroidal velocities of about 10--15\,\kms, in the region probed by the
observations (i.e. 10-50~AU from both the star and jet axis).
 All of these properties, including 
 the implied velocities and angular momentum
 fluxes are in the range predicted by the models, assuming
a central star mass of 0.67~M$_\odot$ (Hartigan et al.~\cite{Hea95}).
The kinematic properties
 of the material surrounding the star are of crucial importance in further
 establishing if the models apply.

DG\,Tauri is known to be surrounded by a circumstellar disk (Beckwith et
al.~\cite{Bea90}; Kitamura et al.~\cite{Kea96a}; Dutrey et al.~\cite{Dea96}).
Previous interferometric observations of the molecular component of
the system (Sargent \& Beckwith~\cite{SB94}; Kitamura et al.~\cite{Kea96b}, 
hereafter KKS)
could not identify a clear signature for rotation around the central object.
These relatively low-resolution (4--5\arcsec), $^{13}$CO(1--0)
observations could not disentangle
the kinematics of the circumstellar disk from the outflow and the outer
envelope velocity fields. In fact, the environment of the star on large
scales appears to be dominated by outflow motions, possibly due to the 
interaction between the outer regions of the disk and the stellar wind
(KKS). In contrast, the inner portion of the disk,
closer to the jet launching region,
is expected to display a Keplerian rotation
pattern. We have carried out new, higher-resolution
millimeter wavelength observations of the \cii\ transition toward the DG~Tauri
system with the aim of
distinguishing the velocity field close to the star and ascertaining
if it is consistent with that expected for Keplerian rotation and 
to check that the disk and jet rotate in the same sense.

\section {Observations and results}


Millimeter wavelength interferometric observations of the DG~Tauri system
was performed using the Owens Valley Radio Observatory (OVRO)
mm-array located near Big Pine, California, between Oct~1999
and Dec~2001.
The six 10.4 meter dishes were deployed in configurations
that provided baselines from 15 to 240~m. Continuum observations centered
at $\sim$220 and $\sim$108~GHz used an analog correlator with a total
bandwidth of 2~GHz. The digital correlator
was configured to observe the \cii\ 
transition with 0.125~MHz resolution over an 8~MHz band 
(0.17 and 11~\kms, respectively).
Frequent observations of 0528$+$134 were used to perform
phase and gain calibration. The passband calibration was obtained by
observing 3C273, 3C454.3 and/or 3C84. The flux density scale was
derived by observing Neptune and/or Uranus, and the calibration
uncertainty is expected to be $\sim 20\%$.
All calibration and data editing used the MMA software package
(Scoville et al.~\cite{mma}). Calibrated ($u,v$) data were then loaded
into the AIPS and/or GILDAS packages for imaging, deconvolution and
analysis. Continuum maps and line cubes were produced using natural
weighting of the ($u,v$) data, and smoothed to a spectral resolution of
0.5~\kms, unless specifically noted. The synthesized beam full width
at half maximum is 1$\farcs$7$\times 1\farcs4$.
Continuum subtraction was performed on the dirty images before 
deconvolution using channels at the edge of the band. 


\subsection{Continuum maps}


We detect unresolved continuum emission from DG~Tauri at both 
1.3 and 2.7~mm. The peak position is the same at both wavelengths,
$\alpha$(2000)$=$04$^{\rm h}$27$^{\rm m}$04.$\!\!^{\rm s}$66
$\delta$(2000)$=$26$^\circ$06$^\prime$16$\farcs$3,
in agreement with previous measurements 
(e.g. Kitamura et al.~\cite{Kea96a}; KKS). The total flux density
is 215~mJy at 222~GHz and 55~mJy at 108~GHz. 
Within calibration uncertainties, the 3 mm value agrees with earlier
interferometer measurements (KKS;
Dutrey et al.~\cite{Dea96}; Looney et al.~\cite{Lea00}) but the 1.3~mm 
value is a factor of two lower than the single dish flux
(Beckwith et al.~\cite{Bea90}), probably because of
spatial filtering by the interferometer. If we assume optically thin emission
from dust grains at T$\sim$40~K and a dust opacity coefficient
k$_\nu$=k$_{\rm 230{\rm GHz}}\times(\nu/230~{\rm GHz})^\beta$, with
k$_{\rm 230{\rm GHz}}$=0.01
(Hildebrand~\cite{H83}, including a gas to dust ratio of 100
by mass), and $\beta\sim$0.5 (Beckwith \& Sargent~\cite{BS91}), our 
measurements imply a total mass of $\sim$0.04~M$_\odot$, consistent
with previous estimates.

\subsection{Line maps}

\begin{figure}
\centerline{\psfig{figure=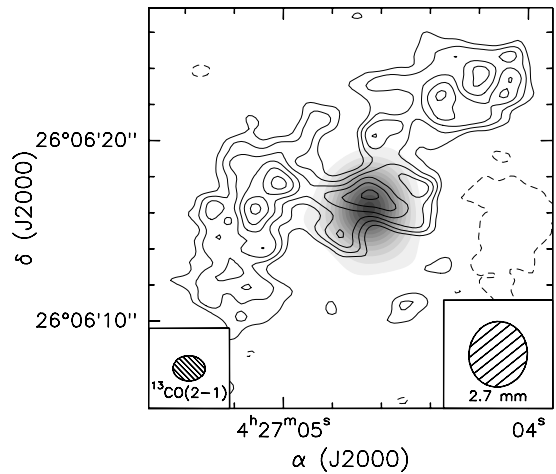,angle=-90,width=7.2cm}}
\caption[]{OVRO \cii\ integrated intensity map (contours) overlaid on the
2.7~mm continuum image (greyscale).}
\label{fcore}
\end{figure}

\begin{figure*}
\centerline{\psfig{figure=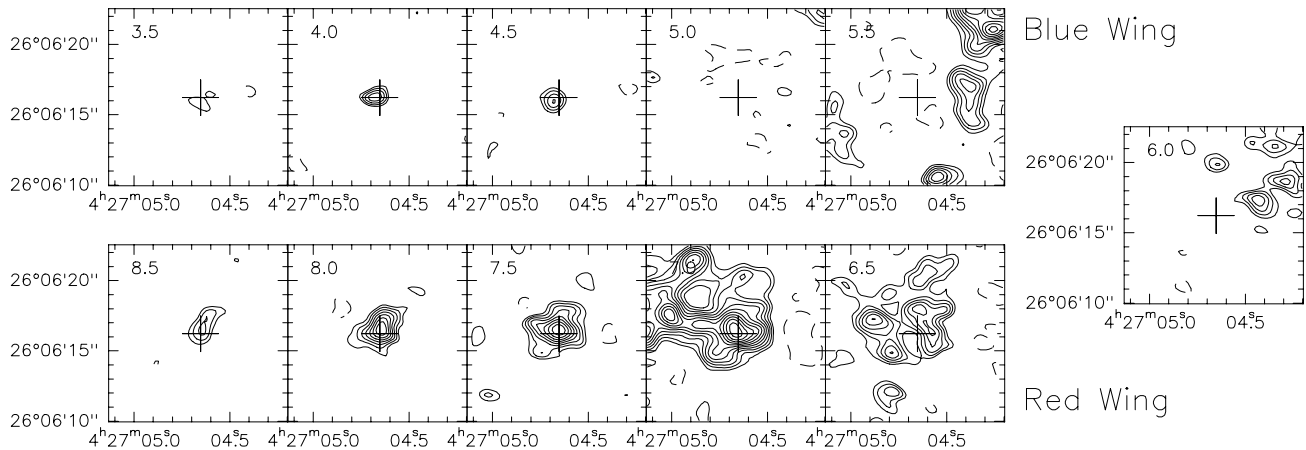,angle=-90,width=17.2cm}}
\caption[]{\cii\ channel maps. Top panels: blue wing; bottom panels: red wing.
Each panel is labelled with the appropriate velocity (v$_{LSR}$ in \kms).
The cross marks the position of the 1.3~mm continuum peak. Contour
levels start at 3$\sigma$ and are spaced by 1$\sigma=60$~mJy/beam.
The central velocity of the system is assumed to be $\sim$6.0~\kms\ (rightmost,
isolated panel).
}
\label{fchan}
\end{figure*}

In Figure~\ref{fcore} we show the \cii\ integrated intensity map overlaid on
the 2.7~mm continuum image. Our observations are sensitive only to the 
compact features of the emission and resolve out most of the extended core
emission seen in the KKS maps. In spite of the filtering
by the interferometer, most of the \cii\ emission is not concentrated
in the inner regions of the system close to the position of the optical star.
The velocity pattern exhibited by this extended gaseous component has been
interpreted by KKS as due to an expanding disk-like
structure, possibly the outer edge of the disk that is being dispersed by 
the stellar wind. Our higher angular resolution map is in good
agreement with their interpretation. In this paper however, 
we will focus on the inner regions of the disk. 

In Figure~\ref{fchan} we show \cii\ channel maps 
(0.5\,\kms\ resolution) of the central
$\sim$12$^{\prime\prime}$ region centered on the continuum peak position
(marked by a cross). The core of the line
(v$_{LSR}$=5.0-7.0~\kms) is dominated by the poorly imaged extended 
structure discussed above. Note that the central velocity channels may
also be affected by self-absorption due to cold foreground
gas (Schuster et al.~\cite{Sea93}).
By contrast, the higher velocity wings, corresponding
to the leftmost three blue and red channels, display 
compact emission arising from the inner disk. The maps
show that the emission peak in the blue channels is shifted toward the 
south-east with respect to the continuum peak, while in the red channels
it is shifted to the north-west. In order to emphasize this velocity gradient,
in Fig.~\ref{fwings} we show the wing emission integrated over the ranges
v$_{LSR}$=3.5--4.5~\kms\ (blue wing) and 7.5--8.5~\kms\ (red wing). These 
maps were obtained with a robust weighting of the ($u,v$) data and the
resulting angular resolution is 1$\farcs4\times$1$\farcs$0.
The red and blue wing emission peaks on opposite sides of the continuum,
which is assumed to trace the stellar position (see also KKS),
and is aligned along a line approximately
perpendicular to the observed direction of the optical jet
(p.a.$\sim$226$^\circ$, marked with a thick line in Fig.~\ref{fwings}).

\section{Discussion}

\begin{figure}
\centerline{\psfig{figure=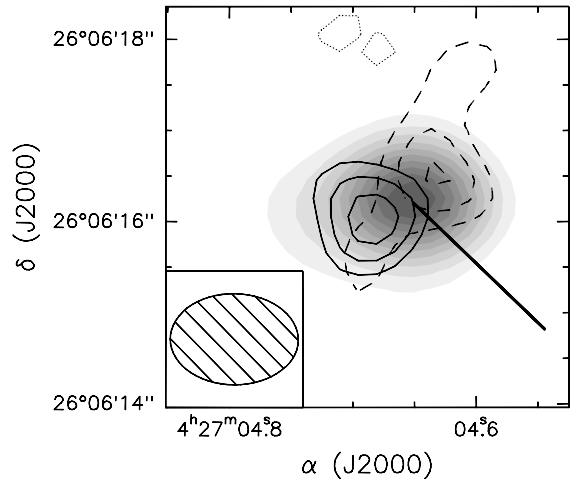,angle=-90,width=7.4cm}}
\caption[]{\cii\ wing maps (contours) overlaid on the 1.3~mm continuum image
(greyscale). Blue and red wing are shown as solid
and dashed contours, respectively.
Contour levels start at 3$\sigma$ and are
spaced by 1$\sigma$, dotted lines show negative contours. The thick solid line
indicates the direction and extent of the initial segment of
the optical jet studied by Bacciotti et al.~(\cite{Bea02}).}
\label{fwings}
\end{figure}

The prime goal of this study was to investigate the velocity pattern
in the inner regions of the DG~Tauri disk and to relate it to the 
velocity pattern detected at the base of the optically visible jet 
by Bacciotti et al.~(\cite{Bea02}). The channel and wing maps of
Figs.~\ref{fchan} and~\ref{fwings} indeed show a velocity
gradient across the inner regions of the circumstellar disk. 
If interpreted as rotation within a disk the axis of which coincides with the
jet axis, the direction of the gradient is consistent with the
sense of rotation inferred for the jet. 

\begin{figure}
\centerline{\psfig{figure=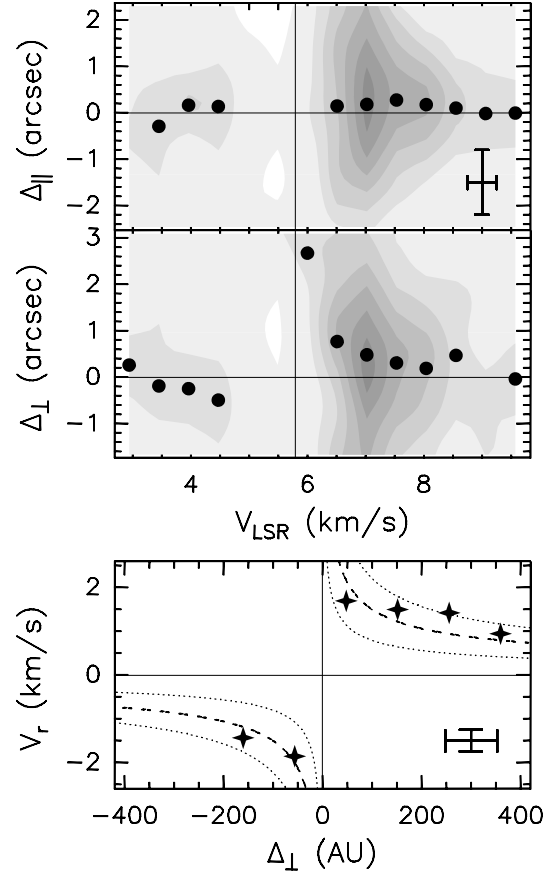,angle=-90,width=7.0cm}}
\caption[]{Top panels: \cii\ position-velocity diagrams parallel (top)
and perpendicular (bottom) to the jet axis.
The crosses
mark the intensity averaged position at each velocity, computed only if the
emission is above 3$\sigma$.
Errorbars in the top panel show the velocity and spatial resolutions.
The thin lines mark the core average velocity
(5.8~\kms) and the position of the continuum source.
Bottom panel: intensity averaged line of sight velocities in the first and
third quadrant of the lower position-velocity diagram,
averaged every 0$\farcs$75 (105~AU).
The dashed line marks the expected line of sight velocity across a Keplerian
disk with the axis inclined by 38$^\circ$ from the line of sight and orbiting
about a 0.67~M$_\odot$ star at a distance of 140~pc from the Sun. The dotted
lines show the range of variation of the Keplerian disk predictions for a
maximum uncertainty of $\pm$15$^\circ$ in the inclination angle and 
$\pm$0.25~M$_\odot$ in the stellar mass.
}
\label{fpv}
\end{figure}

Due to the contamination of the poorly imaged external regions of the disk,
and self-absorption,
it is not possible to study the kinematics of the inner disk
using the line core within 1~\kms\ from the systemic velocity, assumed to
be 5.8~\kms\ (KKS). 
In particular, there can be
no detailed comparison of the observed velocity patterns with Keplerian
rotation models such as those undertaken by
Koerner et al.~(\cite{Kea93}),
Guilloteau \& Dutrey~(\cite{GD94}) and Mannings et al.~(\cite{MKS97}).
Nevertheless,
in Fig.~\ref{fpv} position-velocity diagrams 
along directions parallel and perpendicular to the jet
suggest that, along the disk major axis ($\Delta_\perp$), higher absolute velocity
(with respect to the systemic velocity) peaks are located
closer to the star, while the lower absolute velocities peak 
further away.
Moreover, blue velocities systematically peak
to the south-east and red velocities peak to the north-west of the stellar
position. This behaviour is 
qualitatively consistent with Keplerian rotation in the inner regions of
the disk. A more quantitative comparison with the expected line-of-sight
velocities for a disk surrounding DG~Tauri is shown in
Fig.~\ref{fpv}, bottom panel.
The theoretical curves in the figure were computed
for a central star with mass M$_\star$=0.67~M$_\odot$ (Hartigan et
al.~\cite{Hea95}), and an inclination from the line of sight of
38$^\circ$ (Eisl\"offel \& Mundt~\cite{EM98});
these are the parameters adopted by Bacciotti et al.~(\cite{Bea02})
to check the rotational hypothesis for the jet.
The dotted lines include an uncertainty in these parameters
of $\pm$0.25~M$_\odot$ and $\pm$15$^\circ$.
The observed velocity
pattern in the inner regions of the disk is in excellent agreement
with the model predictions. A more detailed comparison, including the
complete derivation of the disk rotation from molecular line observations,
will require higher angular resolution and more sensitive observations
of optically thinner transitions, such as \ciii, which are possibly less
affected by the external regions of the disk/envelope.

Summarizing our results,
we have shown for the first time that the disk kinematics in a young
T\,Tauri system are in qualitative and
quantitative agreement with the velocity pattern at the base of the jet.
In other words, the simultaneous and kinematically consistent rotation
of disk and jet postulated by the popular magneto-centrifugal
models has been observationally inferred for the first time for the 
region within $\sim$200~AU from the central source.

\begin{acknowledgements}
The OVRO mm array is supported by NSF grant
AST-99-81546, research on young stars and disks
is also supported by the {\it Norris Planetary Origins Project} and
NASA {\it Origins of Solar Systems} program (grant NAG5--9530).
We thank the referee, Chris Davis, for comments that improved the
presentation of our results.
\end{acknowledgements}


\begin{thebibliography}{}
\bibitem[2000]{Bea00} 
 Bacciotti F., Mundt R., Ray T.P., Eisl\"{o}ffel J., Solf J.
 Camenzind M. 2000, ApJ 537, L49
\bibitem[2002]{Bea02} 
 Bacciotti F., Ray T.P., Mundt R., Eisl\"{o}ffel J., Solf J.
 2002, ApJ 576, 222
\bibitem[1990]{Bea90} 
  Beckwith S.V.W., Sargent A.I., Chini R.S., G\"usten R. 1990, AJ 99, 924
\bibitem[1991]{BS91} 
  Beckwith S.V.W. \& Sargent A.I. 1991, ApJ 381, 250
\bibitem[2000]{Dea00} 
  Davis C.J., Berndsen A., Smith M.D., Chrysostomou A., Hobson J.
  2000, MNRAS 314, 241
\bibitem[2000]{Douea00} 
  Dougados C., Cabrit S., Lavalley C., M\'enard F. 2000, A\&A 357, L61
\bibitem[1996]{Dea96} 
  Dutrey A., Guilloteau S., Duvert G., Prato L., Simon M., Schuster K., 
  M\'enard F. 1996, A\&A 309, 493
\bibitem[1998]{EM98} 
  Eisl\"{o}ffel J. \& Mundt R. 1998, AJ 115, 1554
\bibitem[2000]{Eea00} 
  Eisl\"{o}ffel J., Mundt R., Ray T.P. and  Rodr\'{\i}guez L.F., 2000,
  in ``Protostars and Planets IV'', eds Mannings, V.,
  Boss, A.P., Russell, S.S., (Tucson: University of Arizona Press) p. 815
\bibitem[1994]{GD94}
  Guilloteau S. \& Dutrey A. 1994, A\&A 291, L23
\bibitem[1995]{Hea95}
  Hartigan P., Edwards S., Ghandour L. 1995, ApJ 452, 736
\bibitem[1983]{H83}
  Hildebrand R.H. 1983, QJRAS 24, 267
\bibitem[1996a]{Kea96a}
  Kitamura Y., Kawabe R., Saito M. 1996a, ApJ 465, L137
\bibitem[1996b]{Kea96b}
  Kitamura Y., Kawabe R., Saito M. 1996b, ApJ 457, 277 (KKS)
\bibitem[1993]{Kea93}
  Koerner D.W., Sargent A.I., Beckwith S.V.W. 1993, Icarus 106, 2
\bibitem[2000]{KP00} 
  K\"{o}nigl A. \& Pudritz R.E. 2000
  in ``Protostars and Planets IV'', eds Mannings, V.,
  Boss, A.P., Russell, S.S., (Tucson: University of Arizona Press) p. 759
\bibitem[2000]{Lea00}
  Looney L.W., Mundy L.G., Welch W.J. 2000, ApJ 529, 477
\bibitem[1997]{MKS97}
  Mannings V., Koerner D.W., Sargent A.I. 1997, Nature 388, 555
\bibitem[1994]{SB94} 
  Sargent A.I. \& Beckwith S.V.W. 1994, in IAU Coll. n. 140,
  Astronomy with Millimeter and Submillimeter Wave Interferometry, ed.
  M.~Ishiguro \& W.J.~Welch (San Francisco: ASP), 203
\bibitem[2000]{Sea00} 
  Shu F.H., Najita J.R., Shang H., Li, Z.-Y. 2000,
  in ``Protostars and Planets IV'', eds Mannings, V.,
  Boss, A.P., Russell, S.S., (Tucson: University of Arizona Press) p. 789
\bibitem[1993]{Sea93} 
  Schuster K.F., Harris A.I., Anderson N. Russell A.P.G. 1993, ApJ 412, L67
  \& Wang, Z. 1993, PASP 105, 1482
\bibitem[1993]{mma}
  Scoville N.Z., Carlstrom J.E., Chandler C.J., Phillips J.A., Scott S.L.,
  Tilanus R.P.J., \& Wang Z. 1993, PASP 105, 1482
\bibitem[2001]{Wea01}
  Wiseman J., Wootten A., Zinnecker H., McCaughrean M.J. 2001, ApJ 550, L87
\end{thebibliography}
\end{document}